\documentstyle[11pt]{article}
\begin{document}
\begin{titlepage}
\begin{flushright}
SUSSEX-AST 93/6-1 \\
(June 1993)\\
(revised December 1993)\\
\end{flushright}
\begin{center}
\Large
{\bf Extended Inflation with a Curvature-Coupled Inflaton}\\
\vspace{.3in}
\normalsize
\large{Andrew M. Laycock and Andrew R. Liddle} \\
\normalsize
\vspace{.6 cm}
{\em Astronomy Center, \\
School of Mathematical and Physical Sciences, \\
University of Sussex, \\ Brighton BN1 9QH, U.~K.}\\
\vspace{.6 cm}
\end{center}
\baselineskip=24pt
\begin{abstract}
\noindent
We examine extended inflation models enhanced by the addition of a coupling
between the inflaton field and the space-time curvature. We examine two types
of model, where the underlying inflaton potential takes on second-order and
first-order form respectively. One aim is to provide models which satisfy the
solar system constraints on the Brans--Dicke parameter $\omega$. This
constraint has proven very problematic in previous extended inflation models,
and we find circumstances where it can be successfully evaded, though the
constraint must be carefully assessed in our model and can be much stronger
than the usual $\omega > 500$. In the simplest versions of the model, one may
avoid the need to introduce a mass for the Brans--Dicke field in order to
ensure that it takes on the correct value at the present epoch, as seems to be
required in hyperextended inflation. We also briefly discuss aspects of the
formation of topological defects in the inflaton field itself.
\end{abstract}

\begin{center}
\vspace{1cm}
PACS numbers~~~98.80.Cq, 04.50.+h\\
\vspace{1cm}
To appear, {\em Physical Review D}
\end{center}

%%%%%%%%%%%%%%%%%%%%%%%%%%%%%%%%%%%%%%%%%%%%%%%%%%%%%%%%%%%%%%%%%%%%%%
\end{titlepage}
%%%%%%%%%%%%%%%%%%%%%%%%%%%%%%%%%%%%%%%%%%%%%%%%%%%%%%%%%%%%%%%%%%%%%%
\section{Introduction}

Extended inflation \cite{LS,K} has proven to be an extremely interesting
addition to the collection of inflationary universe models \cite{GUTH,KT}, as
it re-introduces the possibility that inflation might proceed via a
first-order phase transition, ending with bubble nucleation, rather than a
second-order phase transition involving the continuous evolution of the
inflaton field into the minimum of its potential. Such a transition is allowed
to complete via the incorporation of extensions to Einstein's gravity, which
alter the space-time dynamics in the presence of a metastable vacuum state.
Such a theory offers a richer phenomenology than traditional `chaotic'
inflationary models \cite{LINDE,KT}, in that the earliest bubbles to nucleate
may be dramatically stretched by the subsequent inflation, providing an extra
source of density perturbations over and above those caused by quantum
fluctuations of scalar fields as present in all inflationary models.

Despite this, extended inflation is not without its problems; indeed, the
problems are closely linked to this bubble production. In original versions
\cite{W,LSB}, the `big bubble problem' simply noted that in the Brans--Dicke
version of the theory, one could only have acceptably small microwave
background perturbations from the bubbles provided the Brans--Dicke parameter
$\omega$ took on a value below around 25, in conflict with present day limits
from solar system time-delay experiments of $\omega > 500$ \cite{RET}. More
detailed calculations \cite{LW} can tighten this number yet further.
Nevertheless, the bubble constraint can easily be evaded by further extensions
of the gravitational sector \cite{AT,LSB,MOD1,MOD2} such as a potential for
the Brans--Dicke field; that the bubble constraint applies during inflation
and the timing experiment at the present epoch leaves plenty of room for
theoretical manoeuvering.

More recently, the big bubble problem has been extended and re-expressed in a
much more restrictive way \cite{LL}. This arises through the recognition that
in order to avoid unacceptably large fluctuations from bubbles, the
modifications to gravity must act to break the scale-invariance of the bubble
distribution. This breaking of the scale invariance inevitably leads to a
breaking of scale-invariance in the density perturbation spectrum {\em from
quantum fluctuations} \cite{KST}, which is taken as requiring the form needed
to explain large scale structure data \cite{LL2}. Recent data, particularly
that from the Cosmic Background Explorer (COBE) satellite \cite{COBE}, has
imposed strong constraints on the extent to which the density perturbation
spectrum can deviate from scale-invariance. The crucial difference in the big
bubble problem expressed in this way is that we now have two constraints
acting in opposite directions {\em at the same cosmological time}. Further,
they appear strong enough to rule out a wide range of extended inflation
models \cite{LL}, including all those known in which inflation is brought to
an end by a first-order transition. We are at present aware of only two
`extended inflation' models in the literature which safely survive these
constraints. One is the Steinhardt--Accetta hyperextended inflation model
\cite{SA,LW2,CS}, in which inflation is brought to an end via some complicated
gravitational dynamics {\em before} any substantial bubble nucleation has
taken place --- the phase transition then completes in the post-inflationary
universe. The second is the `plausible double inflation' model \cite{MOD2}, in
which the first-order phase transition is followed by a second, slow-rolling,
phase of inflation which serves to at least partly erase the bubble
perturbations (for many parameters this slow-roll phase lasts long enough,
more than 70 $e$-foldings, to completely erase the memory of the first-order
inflation). Note that in neither of these models is inflation ended by the
first-order transition.

In this paper, we examine a very simple extension to the basic extended
inflation model which has not been considered in this context before. That is
to include a coupling of the inflaton field to the scalar curvature $R$. This
is a particularly natural term to consider adding to the theory as it does not
involve the introduction of any new fields, and further it has been argued
\cite{LIN82} that in such theories terms of this form are generated by quantum
corrections even if they are not present in the bare (classical) action. Such
a term acts so as to favour the inflaton $\chi$ being trapped in a metastable
$\chi = 0$ state, even if the potential for the inflaton possesses a local
maximum, rather than a minimum, at the origin. In Einstein gravity such a term
is known to be problematical, because if the field is trapped by the curvature
term in such a state the expansion rate tends towards de Sitter in which $R$
is a constant. Such an inflationary state, at least in the classical theory,
persists forever.

Extended inflation offers a crucial difference, because even if the field is
trapped in a metastable state, the solutions governing the expansion do not
tend to a time translation invariant state. For instance, in a Brans--Dicke
theory the scalar curvature falls off as $t^2$ while inflation proceeds.
Consequently, one can arrange a scenario in which the curvature coupling term
initially plays a critical role in aiding inflation, but later becomes
insignificant allowing inflation to end. It is to this possibility that we
devote this paper.

Curvature coupling has arisen in several contexts in the inflation literature.
Our model has connections with all of the following.
\begin{enumerate}
\item Our motivation actually came from a model by Yokoyama \cite{YOK} aiming
to combine cosmic strings with inflation; he used a curvature coupling term
for the string field in order to keep it in its symmetric phase. Inflation was
provided by a slow-rolling field, which naturally gives a decreasing $R$ which
eventually frees the string field.
\item Curvature coupling has been added to chaotic inflation scenarios
\cite{CURV}. In fact, it is subject there to a very strong constraint that the
coupling constant $\xi$ be less than 0.002, otherwise the scenario is ruined.
\item A large negative curvature coupling has been utilized by Salopek, Bond
and Bardeen \cite{SBB} in their `variable Planck mass' model (see also the
earlier work of Spokoiny \cite{SPOK}). As we shall see, their scenario has
several similarities with that discussed here.
\item Two coupled scalar fields were used by Berkin, Maeda and Yokoyama in
their `soft inflation' scenario \cite{SOFT}, and the results obtained were
similar to those of Salopek, Bond and Bardeen \cite{SBB}. The form of their
action allows one field to be interpreted as the Brans--Dicke field via
conformal transformation; however, the fact that the fields in the original
action were coupled means that the second field in the transformed action is
coupled to the Brans--Dicke field through the conformal transformation.
\end{enumerate}

Our aim is to explore the possibilities introduced by this curvature coupling.
The most dramatic new possibility is that inflation can proceed as if driven
by a false vacuum even if the underlying inflaton potential is of second-order
form possessing no metastable state, because of the ability of the curvature
term to stabilize the inflaton at the origin. We shall concentrate most of our
attention on this possibility. In the final section we shall widen our
attention to discuss first-order potentials which already possess metastable
states even without the curvature coupling; this section is closer in spirit
to usual extended inflation scenarios, and the introduction of the curvature
coupling reopens the possibility of ending inflation via bubble nucleation in
extended gravity theories, while evading any constraints from excessively
large bubbles.

\section{The Theory}

\subsection{Equations of motion}

We shall concentrate our entire discussion on the Jordan--Brans--Dicke theory
\cite{BD}, enhanced by a curvature coupling of the inflaton, although
presumably many of these ideas go through in more generally extended models.
The action is given by
\begin{equation}
\label{ACTION}
S = \int {\rm d}^{4} x \sqrt{-g} \left[ -\Phi R + \omega \frac{\partial^{\mu}
	\Phi \partial_{\mu} \Phi}{\Phi} + \frac{1}{2} \partial^{\mu} \chi
	\partial_{\mu} \chi + V(\chi) + \frac{1}{2} \xi R \chi^{2} \right],
\end{equation}
where $\Phi$ is the Brans--Dicke scalar field, $\omega$ the Brans--Dicke
constant, $\chi$ the inflaton field, $R$ the scalar curvature and $\xi$ the
curvature coupling\footnote{One must be wary of the different conventions used
in the literature. With our conventions, conformal coupling occurs for $\xi =
+1/6$.}. We shall not at this stage make a specific choice for the inflaton
potential $V(\chi)$; indeed, we shall not even at this stage state whether it
is of first-order form (metastable vacuum at the origin) or second-order
(local maximum at the origin). Apart from the last term, this is exactly the
action used in the original extended inflation model.

We shall work with the action in this form. However, before proceeding it is
worth looking at the action with the Brans--Dicke field redefined to have the
dimensions and kinetic term of a canonical scalar field, by defining a new
field $\phi^2 = 8 \omega \Phi$ (be careful to distinguish lower and upper
cases). The redefined action becomes
\begin{equation}
\label{SREDEF}
S_{{\rm redef}} = \int {\rm d}^{4} x \sqrt{-g} \left[ \frac{1}{2} \zeta
	\phi^2 R + \frac{1}{2} \partial^{\mu} \phi \partial_{\mu} \phi +
	\frac{1}{2} \partial^{\mu} \chi \partial_{\mu} \chi + V(\chi) +
	\frac{1}{2} \xi R \chi^{2} \right],
\end{equation}
where $\zeta = -1/4\omega$ confirming the Brans--Dicke field as conformally
coupled for $\omega = - 3/2$. We see that the two scalar fields $\chi$ and
$\phi$ appear almost symmetrically in the action, a symmetry which would be
further enhanced by incorporating a potential for $\phi$. Note further that
the action in this form has no pure Einstein--Hilbert term. The different
roles of the two scalar fields, one as inflaton and one as gravitational
scalar, are brought about by the different choices for the potentials (mass
scales of the fields) and by the opposing signs of their curvature couplings.

Returning to the theory as defined by Eq.~(\ref{ACTION}), the field equations
obtained from $S$ by variation with respect to the Brans--Dicke field, the
inflaton field and the metric are
\begin{equation}
\partial^{\mu} \partial_{\mu} \Phi = \frac{\partial^{\mu} \Phi \partial_{\mu}
	\Phi}{2 \Phi} - \frac{ R \Phi}{2 \omega},
\end{equation}
\begin{equation}
\partial^{\mu} \partial_{\mu}\chi = \frac{{\rm d}V(\chi)}{{\rm d}\chi} +
	\xi R \chi,
\end{equation}
\begin{eqnarray}
\left(2 \Phi - \xi \chi^{2} \right) \left( R_{\alpha \beta}-\frac{1}{2}
	R g_{\alpha \beta} \right) & = & \frac{2 \omega}{\Phi} \left(
	\partial_{\alpha} \Phi \partial_{\beta} \Phi - \frac{1}{2}
	g_{\alpha \beta} \partial^{\mu} \Phi \partial_{\mu} \Phi \right)
	- V(\chi) g_{\alpha \beta} \nonumber \\
 & + &  \left(\partial_{\alpha} \chi \partial_{\beta} \chi - \frac{1}{2}
	g_{\alpha \beta} \partial^{\mu} \chi \partial_{\mu} \chi \right)
	+ 2 \left(\partial_{\alpha}\partial_{\beta} \Phi - g_{\alpha \beta}
	\partial^{\mu} \partial_{\mu} \Phi \right) \nonumber \\
 & - &  2 \xi \left( \chi \partial_{\alpha} \partial_{\beta} \chi +
	\partial_{\alpha} \chi \partial_{\beta} \chi - g_{\alpha \beta}
	\left( \chi \partial^{\mu} \partial_{\mu} \chi + \partial^{\mu}
	\chi \partial_{\mu} \chi \right) \right).
\end{eqnarray}

We shall consider these equations in a spatially flat Robertson--Walker
space-time with line element ${\rm d}s^2 = -{\rm d}t^{2}+ a^2(t) {\rm
d}\Omega_{3}^{2}$, where $a(t)$ is of course the scale factor. In accordance
with the assumption of isotropy, we require that our fields have no spatial
dependence. Writing the Hubble parameter as $H = \dot{a}/a$, the equations of
motion become
\begin{equation}
\label{BD}
\ddot{\Phi}+3H\dot{\Phi}=\frac{\dot{\Phi}^{2}}{2\Phi} + \frac{R\Phi}{2\omega},
\end{equation}
\begin{equation}
\label{SCAL}
\ddot{\chi}+3H\dot{\chi}+\xi R\chi = -\frac{{\rm d}V(\chi)}{{\rm d}\chi},
\end{equation}
\begin{equation}
\label{FRIED}
H^{2} = \frac{1}{3 \left( 2 \Phi - \xi \chi^2 \right)} \left[ \omega
	\frac{\dot{\Phi}^2}{\Phi} + \frac{\dot{\chi}^2}{2} + V(\chi)
	- 6 H \dot{\Phi} + 6 \xi H \chi \dot{\chi} \right].
\end{equation}
The standard equations of motion for Einstein gravity with a curvature coupled
scalar (with a sufficiently small $\xi$) can be recovered by setting
$\Phi=m_{Pl}^2/16\pi$, where $m_{Pl}$ is the present day Planck mass, and the
effects of the curvature coupling term can be removed from these equations by
taking their limit as $\xi$ tends to zero. The scalar curvature $R$ is given
by
\begin{equation}
\label{CURV}
R = \frac{-\omega}{\left( 2 \omega + 3 \right) \Phi - \xi \omega \chi^{2}
	\left(1 - 6 \xi \right)} \left[ \left( 2 \omega + 3 \right)
	\frac{\dot{\Phi}^2}{\Phi} + \left( 1 - 6 \xi \right) \dot{\chi}^2
	- 4 V(\chi) + 6 \xi \chi \frac{{\rm d}V(\chi)}{{\rm d}\chi} \right].
\end{equation}

\subsection{Singularities in the equations}

As in the Einstein gravity case, Eqs.~(\ref{FRIED}) and (\ref{CURV}) contain
non-trivial singularities for particular values of the fields.
Eq.~(\ref{FRIED}) contains a singularity when
\begin{equation}
\chi = \chi_1 = \pm \sqrt{\frac{2\Phi}{\xi}},
\end{equation}
and Eq.~(\ref{CURV}) contains a singularity when
\begin{equation}
\chi = \chi_2 = \pm \sqrt{\frac{\left(2\omega+3\right)\Phi}{\left(1 - 6 \xi
\right) \xi \omega}}.
\end{equation}
In the large $\omega$ limit (which we shall be considering
throughout) we have $\chi_1 \simeq \sqrt{1-6 \xi}\, \chi_2$. For $\xi\leq 1/6$
the equation shows us that the singularity in $H$ occurs at a smaller value of
the field $\chi$ than the singularity in $R$. As we shall be considering a
scenario in which the inflaton field evolves from zero to larger values, only
the $H$ singularity need be considered for $\xi\leq 1/6$. It can also be seen
from the above equations that for $\xi \geq 1/6$ that the singularity in $R$
does not exist and we need only consider the singularity in $H$. This shows
that as long as we can avoid the singularity in $H$, the singularity in $R$ is
either also avoided or does not exist. Note that the singularity in the $H$
equation involves the effective Planck mass becoming zero, presumably on route
to negative values, which is certainly something we wish to avoid.

In practice we have shown numerically that the equations are well behaved for
the entire evolution of the system. That this should in fact be the case is
clear in the conformally transformed frame where the potential for the
system rises rapidly as the fields evolve towards those values where the above
singularities arise. The details of this calculation are given in reference
\cite{LAYCOCK}.

\subsection{Other restrictions on the fields}

In addition to the restrictions imposed on the fields in order to avoid the
above singularities, we get further restrictions on the fields due to the
requirement that $H$ and $R$ be both positive and real. The former is required
to have an expanding universe; the latter is not required for dynamical
consistency but is a necessary ingredient for the scenario we shall outline.

Consider first the requirement that $H$ be positive. Eq.~(\ref{FRIED})
expresses $H$ entirely as a function of the fields of the theory. It is easily
shown that
\begin{eqnarray}
\label{EXPAND}
\left( 2 \Phi - \xi \chi^2 \right) H & = & \xi \chi \dot{\chi} - \dot{\Phi}
	\nonumber \\
 & \pm & \frac{1}{3} \left[9 \left(\dot{\Phi} - \xi \chi \dot{\chi}\right)^2
	+ 3 \left( 2 \Phi - \xi \chi^2 \right) \left( \omega
	\frac{\dot{\Phi}^2}{\Phi} + \frac{\dot{\chi}^2}{2} + V(\chi) \right)
	\right]^{1/2}.
\end{eqnarray}
The requirement that $H$ be positive obviously leads us to the use of the
positive root in our calculations and we also get a restriction on the field
values due to the requirement that the argument of the root in
Eq.~(\ref{EXPAND}) be greater than zero. However, this is easily seen to be
satisfied as long as $\chi \leq \chi_1$ (assuming only that the Brans--Dicke
field is positive), and hence provides no additional restrictions.

The model that we are considering requires us to keep $R$ positive, at least
in the initial stages of the evolution of the system. Looking at
Eq.~(\ref{CURV}) it can be seen that if we take $\omega$ positive (which we
always do) then the prefactor to the square bracket will always be negative if
$\chi \leq \chi_2$. As this condition will always be satisfied (as we require
our system to evolve in such a way as to avoid the singularities in the
equations) the requirement that $R$ be positive becomes
\begin{equation}
\label{INEQ}
\left( 2 \omega + 3 \right) \frac{\dot{\Phi}^2}{\Phi} + \left( 1 - 6 \xi
	\right) \dot{\chi}^2 - 4V(\chi) + 6 \xi \chi
	\frac{{\rm d}V(\chi)}{{\rm d}\chi} \leq 0.
\end{equation}
In general it is difficult to say whether or not this condition is satisfied.
However, in the scenarios outlined below it can be shown that in what we shall
refer to as the trapped phase of the system this inequality becomes a
restriction on the Brans--Dicke constant $\omega \geq - 1/2$. It will also be
seen that in the case of the second-order potential that, for the values of
the parameters which we have chosen to meet the various constraints on the
system, the system evolves in such a way as to ensure that this condition will
always be satisfied. Numerical simulation of the system's evolution has also
shown that for the parameters chosen the inequality Eq.~(\ref{INEQ}) is always
satisfied.

\section{Curvature-Coupled Extended Inflation}

\subsection{Initial conditions}

We shall consider a scenario in which the inflaton field is initially
trapped, with the aid of its curvature coupling, at $\chi=0$. As we shall see,
it depends on parameters whether or not the trapped phase is actually relevant
for details such as density perturbation production. For definiteness, let us
assume for the time being that the underlying potential $V(\chi)$ is of the
simplest form for a second-order phase transition, that is
\begin{equation}
\label{2ND}
V(\chi) = \frac{\lambda}{4} \left( \chi^2 - \chi_0^2 \right)^2,
\end{equation}
where $\lambda$ and $\chi_0$ are constants\footnote{After submitting our
paper, we received a preprint by Linde \cite{LINP} which, albeit with a
different emphasis, discusses the model given by this choice of potential. We
are extremely grateful for many subsequent discussions with the author, which
have led to several important improvements to this paper.}. Many results will
depend only on the energy of the metastable state, and where that is the case
we shall denote it by $M^4$. For this potential $M^4 = \lambda \chi_0^4/4$.

Ordinarily this potential wouldn't admit metastable states, as $\chi=0$ is a
local maximum of the potential. The introduction of the curvature coupling
changes this. We shall often refer loosely to an `effective potential'
\begin{equation}
\label{EFFPOT}
V_{{\rm eff}}(\chi) = V(\chi) + \frac{1}{2} \xi R\chi^{2},
\end{equation}
though we stress that as $R$ depends on all the fields of the theory this
terminology is not strictly accurate. However, the equations of motion show
that it applies provided $R$ is slowly varying, and it is very useful for
building a picture of the possible dynamics. By analogy to the
temperature-corrected effective potential, provided $R$ is sufficiently large
the field can be trapped at $\chi = 0$. For the potential above, one requires
\begin{equation}
R > R_{c} = \frac{\lambda\chi_{0}^{2}}{\xi}.
\end{equation}
As long as $R$ remains greater than $R_c$, $\chi=0$ remains the appropriate
stable solution to the classical equations. As noted in the introduction, in
general relativity the corresponding solution for $R$ is that it remains
constant, and so classically the solution pertains forever.

We pause briefly to comment on how the field may have got trapped in the first
place. There is an important difference from a temperature correction in that
$R$ depends on $\chi$. Indeed, it is the potential energy of the $\chi$ field
that provides the dominant contribution to $R$. Ordinarily there is another
perfectly good minimum of the energy functional where $\chi$ sits close to its
classical minimum $\chi_0$, and $R$ is very small with only the Brans--Dicke
kinetic contribution. Indeed, presumably it is possible, even in general
relativity, for the field to tunnel via a form of gravitational instanton into
this second minimum. This highlights the dangers of thinking of
Eq.~(\ref{EFFPOT}) as an effective potential, as this possibility would be very
hard to see. Although tunnelling by gravitational instanton provides some
interesting possibilities, they are beyond the scope of this paper and we
shall assume from here on that the probability of such tunnelling is
negligibly small.

Despite having set aside tunnelling, one must still address why one shouldn't
expect the field to originally settle in the true vacuum state. This is easily
resolved --- it seems likely that the field can become trapped in the
metastable $\chi=0$ state by thermal effects, exactly as envisaged in any
first-order inflation model. Certainly one expects the gravitational term to
assist, rather than interfere, with this process. Having deposited the field
close to the $\chi=0$ state, the thermal effects can then fall off as
inflationary supercooling takes place, leaving the curvature term to hold the
field in the metastable state. [Later we shall also consider the possibility
that the bare potential itself has a metastable state.]

\subsection{Approximations to inflationary evolution}

Before embarking on a study of the exact solutions in various regimes, we
shall here derive an approximate relationship between the scale factor and the
Brans--Dicke field during inflation which shall prove very pertinent later.
This result is geometric in nature, and does not depend on the details of the
fields driving inflation. Suppose that inflation is proceeding with $a(t)
\propto t^p$ for some high exponent $p$. As $R$ has the geometrical definition
\begin{equation}
R = 6 \left( \frac{\ddot{a}}{a} + \left( \frac{\dot{a}}{a} \right)^2 \right)
	\,,
\end{equation}
one finds
\begin{equation}
R = \left(12-\frac{6}{p} \right) H^2 \simeq 12 H^2 \,.
\end{equation}
That is, as long as inflation is proceeding sufficiently rapidly the scalar
curvature is given simply by $12 H^2$. Taking this to be the case, and
substituting this into the equation for $\Phi$, dropping the $\ddot{\Phi}$
term by use of a slow-roll approximation\footnote{One might suggest that
instead one should define a canonical scalar field such as $\phi$ used in
Eq.~(\ref{SREDEF}) and drop the $\ddot{\phi}$ term; this alters the
coefficient of the second term in Eq.~(\ref{HPHI}) but does not change the
approximate solution Eq.~(\ref{HPHISOL}).}, yields an algebraic relation
between $\dot{\Phi}/\Phi$ and $H$
\begin{equation}
\label{HPHI}
6 H \frac{\dot{\Phi}}{\Phi} - \frac{\dot{\Phi}^2}{\Phi^2} - \frac{12}{\omega}
	\, H^2 = 0 \,,
\end{equation}
the appropriate solution to which is
\begin{equation}
\label{HPHISOL}
H \simeq \frac{\omega}{2} \frac{\dot{\Phi}}{\Phi} \,.
\end{equation}
Integrating yields
\begin{equation}
\label{GROWTH}
a(t) \propto \Phi^{\omega/2} (t) \,.
\end{equation}

Although approximate, this result is useful because it applies regardless of
whether inflation is driven by a trapped or rolling field in a Brans--Dicke
theory. Further, we have found numerically that it provides an excellent
approximation, giving an analytic link between the number of $e$-foldings of
inflation and the growth of the Brans--Dicke field.

\subsection{The trapped phase}

While in the trapped phase, the solution is extremely simple, because the
equations reduce directly to the standard extended inflation ones \cite{N,LS}.
Along with $\chi(t) = 0$, one has
\begin{eqnarray}
\Phi(t) & = & \Phi_0 \left( 1 + B t \right)^2,\\
a(t) & = & a_0 \left( 1 + B t \right)^{\omega + 1/2},
\end{eqnarray}
where $B^2=M^4/6\alpha^2\Phi_0$ and $\alpha^2 = \left(2 \omega + 3 \right)
\left( 6 \omega + 5 \right)/12$. This is not the most general solution, but it
is known to be a late-time attractor. It almost satisfies the approximate
relation Eq.~(\ref{GROWTH}), with a small correction accounting for the
avoidance of a slow-roll approximation. One can readily calculate the behaviour
of the scalar curvature to be
\begin{equation}
R(t) = \frac{6 \omega \left( 2 \omega + 1 \right) B^2 \Phi_0}{\Phi(t)} =
	\frac{6 \omega \left( 2 \omega + 1 \right)}{\left( 2 \omega + 3
	\right) \left( 6 \omega + 5 \right)} \, \frac{2M^4}{\Phi(t)},
\end{equation}
where the prefactor on the right hand side is very close to unity. As
advertized, the scalar curvature falls as $t^{2}$ while in the trapped phase.
It is thus inevitably destined to pass below $R_c$, bringing the trapped phase
to an end. Regardless of initial conditions, this occurs at the same value of
the Brans--Dicke field --- for the second-order potential this is at
\begin{equation}
\Phi_c = \frac{6\omega\left(2\omega+1\right)}
	{\left(2\omega+3\right)\left(6\omega+5\right)} \,
	\frac{\xi \chi_0^2}{2}.
\end{equation}
Sufficient inflation is arranged by choosing a small enough initial value of
$\Phi$. However, one must be careful to avoid breaching the classical regime.
Comparing the action with the Einstein--Hilbert one, the effective Planck mass
at a given epoch during the trapped phase is $m_{Pl}^{{\rm eff}} = \sqrt{16\pi
\Phi}$, requiring $\Phi_0 \gg M^2/16\pi$.

\subsection{The rolling phase}

With this second-order potential, once $R < R_c$ it becomes favourable for the
field to roll down the potential towards the minimum at $\chi= \chi_0$. It
will be displaced from the local maximum of the potential by quantum
fluctuations. When the inflaton field begins to evolve, the simple solutions
of the trapped phase no longer apply and an analytic solution to the equations
of motion is no longer possible. A very important feature of this rolling is
that the Planck mass begins to {\em decrease}, as the curvature coupling of
the $\chi$ field makes its influence felt. As we shall see, this is very
significant.

In general it is possible for a significant amount of inflation to occur
during this phase. The amount which occurs will depend on all the parameters
of the theory, and we can identify two cases with distinct physics. The first
is that the number of $e$-foldings in the rolling phase is substantially less
than sixty. In such a case, the field was still in the trapped phase when
scales corresponding to the present observable universe crossed the Hubble
radius during inflation, and particularly significantly this means that the
density perturbations leading to large scale structure formation were
generated during the trapped phase. The second case corresponds to having more
than sixty $e$-foldings in the rolling phase. In that case, none of the
physics pertaining to the presently observable universe is linked directly to
the physics of the trapped phase. Nevertheless, even in this case the prior
existence of the trapped phase is assumed, leading to a precise form of
initial conditions for the much more complex rolling phase.

Once the field reaches its minimum, it will oscillate and give a conventional
reheating scenario. There is unlikely to be much difference here from
conventional treatments, and in any case the advent of at least the
possibility of electro-weak baryogenesis \cite{DOLGOV} has effectively removed
any stringent reheating constraint anyway. What is much more significant in
this phase is the behaviour of the Brans--Dicke field, as we shall shortly
discuss.

In the limit of small $\xi$ ({\it ie} where the inflaton is decoupled from the
curvature) it has been shown by Linde \cite{Lin90} that the relationship
between the scale factor and the Brans-Dicke field $\Phi$ given by
Eq.~(\ref{GROWTH}) holds even in the rolling phase. The analytic arguments of
section III.B indicate that it should be a good approximation even at large
$\xi$ and we have verified this numerically. This has some very important
consequences for this theory which are discussed below.

\section{Parameter Constraints}

Our model is rich in parameters. Even ignoring initial conditions (which it is
assumed are rendered irrelevant by the trapped phase inflationary attractor),
one has $\omega$, $\xi$ and at least two parameters describing the inflaton
potential. However, there are substantial constraints that must be obeyed, and
in other cases there are at least guideline values. In the latter category is
the value of the symmetry breaking scale for the inflaton potential, $\chi_0$.
Primarily we shall consider $\chi_0$ to be a typical GUT scale parameter in
the range $10^{-5} m_{Pl}$ to $10^{-3} m_{Pl}$, though in complete generality
its range of plausible values is much wider.

Let us progress through the constraints, assuming the second-order potential
of Eq.~(\ref{2ND}).

\subsection*{Classical behaviour and sufficient inflation}

We have already remarked above that classicality demands $\Phi_0 > M^2/16\pi$.
When we couple this to the requirement that there be a trapped phase, $\Phi_0
< \Phi_c$, this requires $\sqrt{\lambda} < 8\pi \xi$. This can be further
strengthened by the requirement of sufficient inflation, the usual requirement
being a minimum of seventy $e$-foldings to ensure the solution of the usual
cosmological problems. The number $N$ of $e$-foldings required in the trapped
phase depends on the number available in the rolling phase; once this has been
calculated the trapped phase solutions yield $\Phi_0 < \exp \left(
-4N/(2\omega+1) \right) \Phi_c$ tightening the constraint on $\lambda$
somewhat.

\subsection*{Normalization of the present day Planck mass}

This is the most crucial constraint on the scenario. One must reproduce the
present day value of the Planck mass, which with our conventions requires that
the coefficient of $R$ be $- m_{Pl}^2/16\pi$. By the present the $\chi$ field
will have fallen into its true vacuum state, so the requirement is
\begin{equation}
\Phi_p-\xi \chi_0^2/2 = m_{Pl}^2/16\pi,
\end{equation}
where $\Phi_p$ is the present day value. We see a crucial aspect of the
scenario here; with this choice of potential, were $\Phi$ to retain its
critical value then the effective Planck mass would be cancelled almost
exactly between the two fields! Hence the present day Planck mass is entirely
given by the growth in $\Phi$ since the critical point. It is well known that
for large $\omega$ the fractional growth in $\Phi$ during conventional
radiation or matter domination is small \cite{N,W} if $\Phi_c$ is a free
field. Hence without a potential the only phase during which $\Phi$ may grow
significantly beyond $\Phi_c$ is during the rolling phase. A precise
calculation of this amount of growth requires numerical study, but the
analytic relationship Eq.~(\ref{GROWTH}) provides an excellent indication.

Let us parametrize this growth, whether its source be dynamical or a
potential, via a quantity $k$, which in general depends on all the parameters
in the action, by writing
\begin{equation}
\Phi_p = k \Phi_c.
\end{equation}
Then the normalization condition for this potential requires
\begin{equation}
\label{NORM}
(k-1) \Phi_c = m_{Pl}^2/16\pi,
\end{equation}
which for the second-order potential implies that one must have
\begin{equation}
\label{KEQ}
k=1+\frac{\left(2\omega+3\right)\left(6\omega+5\right)m_{Pl}^2}
	{6\omega\left(2\omega+1\right)8\pi\xi\chi_0^2} =
	1+f\left(\omega,\xi,\chi_0\right),
\end{equation}
if the Brans--Dicke field is to be correctly normalized. In general, the
evolution of $\Phi$ will lead to a $k$ value that does not satisfy this
constraint, symptomizing a failure to match the present day Planck mass. The
parameters must be selected in order that this condition is met.

As an excellent guide, Eq.~(\ref{GROWTH}) gives an expression for $k$ in terms
of the number of $e$-foldings $N$ that occur during the rolling phase
\begin{equation}
\label{KINN}
N = \frac{\omega}{2} \ln k \,.
\end{equation}

If the fractional growth of the field after the trapped phase is small, then
the critical value of $\Phi$ is larger than would be required to give the
present day Planck mass because of the counteracting effect of the $\chi$
field; $k$ will be less than 2 in Eq.~(\ref{NORM}). That is, the effective
Planck mass {\em increases} up to the critical point, and thereafter {\em
decreases} to its present day value.

We shall find that the near cancellation of the two contributions is an
obstacle to model building. Our discussion will follow three possibilities, of
roughly decreasing attractiveness.
\begin{enumerate}
\item The present day Planck mass is determined by the dynamical growth of
$\Phi$ after passing the critical point.
\item The present day Planck mass is determined by the introduction of a
potential for $\Phi$ as utilized in hyperextended inflation \cite{CS}. This
potential will have a minimum at $\Phi = m_{Pl}^2/16 \pi + \xi \chi_0^2/2$,
which is generically at a larger value of $\Phi$ than the critical value. It
must be acknowledged though that such a potential involves very serious
fine-tuning if it is not to interfere with the inflationary dynamics
\cite{AFG}.
\item The cancellation between the two scalar field terms is a
consequence of the specific potential chosen, as it is caused by a precise
relation between the curvature at the top of the bare potential $V(\chi)$ and
the vacuum expectation of $\chi$ in the true vacuum. For more general
potentials such a cancellation need not occur, and the factor $(k-1)$ would be
unlikely to be much less than unity. However, the potential as given is the
most general renormalizable potential with a $\chi \rightarrow - \chi$
symmetry and zero potential energy in the true vacuum state, and one would
need strong motivation to go to a more general case.
\end{enumerate}

\subsection*{Large enough $\omega$}

In a pure Brans--Dicke theory, the present day limit on $\omega$ is that it
should exceed 500 \cite{RET}, though there are no constraints if the
Brans--Dicke field is anchored by a potential. Our model is more complex, due
to the competing contributions to the present gravitational `constant', and so
one must be careful in assessing the constraint.

In the literature \cite{PNP}, constraints are quoted via post-Newtonian
parameters derived from a gravitational action written in the form
\begin{equation}
S_{{\rm grav}} = \int {\rm d}^4 x \sqrt{-g} \left[ - \tilde{\Phi} R +
	\tilde{\omega}(\tilde{\Phi}) \frac{\partial^{\mu} \tilde{\Phi}
	\partial_{\mu} \tilde{\Phi}}{\tilde{\Phi}} \right],
\end{equation}
which is that of a general scalar-tensor theory. At the present it is fair to
assume that the $\chi$ field is fixed in its vacuum, and so the relevant
transformation is
\begin{equation}
\tilde{\Phi} = \Phi-\xi \chi_0^2/2 \,,
\end{equation}
yielding
\begin{equation}
\tilde{\omega}(\tilde{\Phi}) = \omega \frac{\tilde{\Phi}}{\tilde{\Phi}+
	\xi \chi_0^2/2} \,.
\end{equation}
The post-Newtonian parameters depend on $\tilde{\omega}$ and ${\rm d}
\tilde{\omega}/{\rm d}\tilde{\Phi}$, but for this $\tilde{\omega}$ one can
ignore the latter (which gives a $1/\tilde{\omega}$ correction) and take the
constraint as $\tilde{\omega} > 500$ \cite{PNP}. Using the $k$-parametrization
of the above subsection then yields
\begin{equation}
\tilde{\omega}=\frac{\left(k-1\right)\omega}{k}
\end{equation}

Any system we propose must yield a value of $\tilde{\omega}$ greater than 500.
As long as the cancellation between the two contributions to the Planck mass
is not very precise, this gives the standard result. But, as we have
commented, the fractional growth of $\Phi$ from the critical point can be very
small, for some parameters only a few percent. In that case, the solar system
constraint is much stronger than expected. The physical reason for this is
clear; the larger the $\Phi$ field's value, the more dramatic the effects of
its variation and the larger $\omega$ has to be to damp them down.

Let us finally remind ourselves that in the case where there is also a
potential for $\Phi$, there is no $\omega$ constraint as the potential anchors
the field, preventing the (spatial) variation which gives rise to the
deviations from the general relativistic predictions in our solar system.

\subsection*{Density perturbations}

The strongest argument in favour of inflation is its ability to produce
adiabatic density perturbations, and it is reasonable to require that these be
of the appropriate size to match observations. We take this constraint as
requiring that the COBE $10^{\circ}$ microwave anisotropy is reproduced. The
perturbations on interesting scales are produced around 60 $e$-foldings from
the end of inflation, and there are two possible situations that may arise.
\begin{enumerate}
\item {\bf The $\chi$ field is in the trapped phase 60 $e$-foldings from the
end of inflation.}\\
In this case the calculation is just that of the standard extended inflation
scenario \cite{KST}. The standard technique \cite{SBB} for calculating density
perturbations in scalar-tensor theories is the conformal transformation, by
which the gravitational term is brought into the Einstein--Hilbert form,
followed by field redefinitions to render the Brans--Dicke kinetic term into
canonical form. In the trapped phase we can assume that the inflaton degrees
of freedom are negligible, and concentrate only on the Brans--Dicke field.
Utilizing a conformal factor so that $\tilde{g}_{\mu \nu} =
(16\pi\Phi/m_{Pl}^2) g_{\mu \nu}$, this yields a transformed action
\begin{equation}
S_{{\rm con}} = \int d^{4}x \sqrt{-\tilde{g}} \left[ -\frac{m_{Pl}^{2}\tilde{R}
          }{16\pi} + \frac{1}{2} \partial_{\mu}\psi \partial^{\mu}\psi + \exp
	  \left( -\frac{2\psi}{\psi_{0}} \right) M^{4} \right],
\end{equation}
where the field $\psi$ is defined as follows
\begin{eqnarray}
\psi & = & \psi_{0}\ln\left(\frac{16\pi\Phi}{m_{Pl}^{2}}\right), \\
\psi_{0}^{2} & = & \frac{\left(2\omega+3\right)m_{Pl}^{2}}{16\pi}.
\end{eqnarray}

One can then apply the standard formula for calculating the perturbation
amplitude. The final step is to assume that at late times the conformally
transformed frame and the untransformed frame coincide to high accuracy, so
that the result applies in the original frame. Although this appears largely
motivated by a laziness regarding transforming back, the known accurate
constancy of the gravitational coupling over the last few Hubble times
effectively guarantees it. The density perturbation $\delta_H$ (effectively
the density contrast at horizon crossing and formally defined as in
\cite{LL2}) is given by
\begin{equation}
\label{PERT}
\delta_H = \sqrt{\frac{512\pi}{75}} \, \frac{V^{3/2}(\psi)}{m_{Pl}^3 \left(
	{\rm d} V(\psi)/{\rm d}\psi \right)} = \frac{1}{20\pi} \left[
	\frac{\left( 2 \omega + 3 \right)}{6} \right]^{1/2} \left(
	\frac{M}{m_{Pl}} \right)^2 \frac{m_{Pl}^2}{\Phi},
\end{equation}
where the right hand side is evaluated at horizon crossing $\kappa = aH$,
where $\kappa$ is the comoving wavenumber. The COBE result \cite{COBE}
requires\footnote{This value is appropriate to nearly flat spectra with
negligible gravitational waves, which is exactly the situation here as in the
transformed frame the model corresponds to power-law inflation with a very
high exponent.} that $\delta_H$ evaluated 60 $e$-foldings from the end of
inflation equals $1.7 \times 10^{-5}$ \cite{LL2}. The equation for $\delta_H$
can be evaluated explicitly 60 $e$-foldings from the end of inflation if we
make the approximation that $\Phi$ at the end of inflation is equal to
$\Phi_{c}$, the value of $\Phi$ at the critical point. One can then use the
trapped phase solutions to evaluate $\Phi$ 60 $e$-foldings from the end of
inflation. This actually gives us a slight overestimate of the size of
$\delta_H$, but for the cases we discuss the correction (which appears in the
exponential term below) is at most a few percent. This gives us a value for
$\delta_H$ of
\begin{equation}
\label{DENSPER}
\delta_H = \frac{F\left(\omega\right)}{20\sqrt{3} \, \pi} \,
	\sqrt{\frac{\omega \lambda}{\xi^2}} ,
\end{equation}
where $F\left(\omega\right)$ is given by
\begin{equation}
F\left(\omega\right) = \left(\frac{2\omega+3}{2\omega} \right)^{\frac{1}{2}}
	\frac{\left(6\omega+5\right)\left(2\omega+3\right)}{6\omega
	\left(2\omega+1\right)}\exp \left(\frac{240}{2\omega+1}\right),
\end{equation}
and is close to unity for large $\omega$. If $N_{{\rm roll}}$ $e$-foldings
occur during the rolling phase, the term in the exponential is modified to
$(240-4N_{{\rm roll}})/(2\omega+1)$, but for typical parameters (primarily
large $\omega$) this is a tiny correction, saving one the trouble of the
numerical calculation of $N_{{\rm roll}}$.

\item {\bf The $\chi$ field is in the rolling phase 60 $e$-foldings from the
end of inflation.}\\
This is a much trickier situation. Because both scalar fields are evolving,
and both couple to the curvature tensor, the simple conformal transformation
above cannot be used. We do not intend to investigate this situation here due
to this complexity, though unfortunately this will restrict our ability to
make precise statements in this case. A general formalism for dealing
numerically with multiple non-minimally coupled fields is provided elsewhere
\cite{LAYCOCK}, and would be applicable in more general circumstances,
including the small $\xi$ limit of the theory as investigated by Linde
\cite{LINP}.
\end{enumerate}

\subsection*{Putting it all together}

With a free Brans--Dicke field, it proves impossible to satisfy all the
constraints if one demands that large scale structure perturbations were
generated during the trapped phase. This only requires the solar system
constraint and the approximate relation between $N$ and $k$; one has
\begin{equation}
N  \simeq \frac{\omega}{2} \ln k > 250 \frac{k}{k-1} \, \ln k
\end{equation}
By definition $k > 1$, which implies that the $k$-dependent factor always
exceeds unity and hence the number of $e$-foldings in the rolling phase always
greatly exceeds 60 when the solar system bounds are satisfied (note that this
hasn't even required the correct Planck mass normalization). The trapped phase
can therefore only find its role in providing unique initial conditions for
the rolling phase.

By consequence, the scenario is substantially complicated in that the density
perturbation calculation fits into the second case, where analytic progress is
not possible. Nevertheless, there is no reason to suppose that a satisfactory
level of density perturbations cannot be achieved, as the parameter $\lambda$
is not fixed by any other criterion and can therefore be selected to fit the
perturbation constraint. This is because we find numerically, but have been
unable to confirm analytically, that the amount of inflation and the growth of
$\Phi$ are independent of $\lambda$, as is the case in general relativity.
Consequently the details of the dynamical normalization are unaffected by
adjusting $\lambda$ to ensure the correct level of perturbations. A detailed
investigation of the perturbation constraint must however be reserved for
future work.

Setting aside the perturbation constraint, it is clear that the correct
present day Planck mass can be achieved, as Eq.~(\ref{KEQ}) amounts to a
single constraint on the $\xi$--$\chi_0$--$\omega$ parameter space. This
constraint yields a two-dimensional surface in the parameter space, which must
be calculated numerically. For every point on that surface there is then
associated a value of the parameter $\tilde{\omega}$ which may or may not
satisfy the solar system timing constraint; this too can only be determined by
use of numerical calculations. We have shown numerically that it is indeed
possible to find parameters satisfying all the constraints, though the
parameter space is too extensive for us to have investigated its limits. For
example, the correct present day value of the Planck mass is obtained if we
choose the parameters to be $\chi_0=10^{-3}$, $\xi=7 \times 10^5$ and
$\omega=1000$. This particular choice of parameters does not however result in
a model which satisfies the constraint on the parameter $\tilde{\omega}$; it
gives a value of $\tilde{\omega}=53$.

Working from this example, it is possible to obtain a picture of the two
dimensional surface in the $\xi$--$\chi_0$--$\omega$ parameter space. First,
further working models can be obtained by changing the values of the
parameters $\xi$ and $\chi_0$ but keeping the product $\xi\chi_0^2$ constant,
{\it ie} $\chi_0=10^{-5}$, $\xi=7 \times 10^9$ and $\omega=1000$ produces a
model with exactly the same values for the Planck mass and $\tilde{\omega}$.

Secondly, it is possible to obtain further models which produce the correct
value of the Planck mass, by changing $\xi$ and $\chi_0$ in  such a way that
the product $\xi\chi_0^2$ does not remain constant, but then we must also use
a different value for $\omega$. For example $\chi_0=10^{-3}$, $\xi=7 \times
10^6$ and $\omega=10^5$. In this case $\tilde{\omega}=546$, satisfying the
constraint on this parameter. Of course in this case the system is in the
rolling phase 60 $e$-foldings from the end of inflation, as the calculation
above proves it must be. Again, equivalent models, with the same value for
$\omega$ and $\tilde{\omega}$, can be obtained by varying $\xi$ and $\chi_0$
but keeping their product constant.

If the use of a free Brans--Dicke field is abandoned, then models become much
easier to construct, as the potential invalidates both the $\omega$ constraint
and the need to normalize the Brans--Dicke field dynamically. It is assumed
that such fine-tuning has been introduced \cite{AFG} as to make the
Brans--Dicke potential flat enough during inflation to have negligible affect
on the dynamics.

By allowing $\tilde{\omega}$ to be less than 500, we can obtain working models
with a wide range of values of the parameters $\omega$, $\chi_0$ and $\xi$. In
particular, it becomes possible, through a sufficiently low choice of
$\omega$, that the inflaton is still in the trapped phase 60 $e$-foldings from
the end of inflation, as was the case for the first example quoted above. Then
the simpler density perturbation calculation applies and can be used directly
to fix the value of $\lambda$; the possible range of values for $\lambda$ is
also wide. For the first example set of parameters quoted above, $\lambda$ is
of order $10^{3}$.

\section{Curvature Coupling and First-Order Inflation}

In the sense that the relevant evolution is completely classical, we have up
until now been discussing what is essentially a slow-rolling inflation model,
with the novelty of the inflaton field being trapped in a false vacuum state
for a finite duration by the curvature coupling. In this section we consider a
case closer to the heart of the original extended inflation model, by
considering an underlying inflaton potential $V(\chi)$ which itself exhibits a
metastable vacuum state in which the field is trapped. Such a situation may
arise for example via a Coleman--Weinberg type of potential \cite{CW}. For
simplicity we model this potential via a quartic polynomial
\begin{equation}
V(\chi) = M^4 + \beta \chi^2 + \gamma \chi^3 + \delta \chi^4,
\end{equation}
which will allow us to utilize tunnelling results found by Adams \cite{ADAMS}.
An appropriate choice of the parameters gives a metastable vacuum at the
origin and a true vacuum state on the positive $\chi$ side. Although this
potential does not have the symmetry about $\chi=0$ that would ensure full
consistency with our incorporation of the curvature term, we expect that the
effect of the curvature coupling term would be the same for a wide range of
first-order potentials. This particular potential has been chosen as a simple
illustration of the effect of the curvature coupling.

Ordinarily in extended inflation this theory suffers from the big bubble
problem, because the rate of efficiency of nucleation is only a modest function
of time during inflation. Nucleation efficiency is normally expressed via the
dimensionless parameter describing the nucleation per hubble volume per hubble
time
\begin{equation}
\epsilon = \Gamma/H^4,
\end{equation}
where $\Gamma$ is the nucleation per unit volume per unit time. For an
unchanging potential $\Gamma$ is constant. The criteria for the phase
transition to complete is that $\epsilon$ be of order one, which is brought
about in extended inflation by the time variation of $H$. However, $\epsilon$
grows slowly by comparison with the scale factor, so bubble nucleation can
still be sufficiently probable in the earlier stages (say 50 $e$-foldings from
the end) to allow large bubbles to be generated.

Our aim is to use the curvature coupling to suppress this early nucleation.
One can see the motivation from the `effective potential'. When $R$ is
sufficiently large, it is the $\chi = 0$ state which is the minimum, and there
can be no tunnelling, barring the possibility of gravitational instantons as
discussed earlier\footnote{Of course, further work is required to decide
whether or not this possibility really can be consistently ignored.}. Only as
$R$ drops does the `effective potential' develop the second minimum. Once it
falls below the $\chi=0$ minimum, nucleation then becomes favourable,
increasingly so as $R$ falls. Ideally, the underlying potential should possess
a very rapid nucleation rate $\epsilon$, so that once the effect of the
curvature coupling term is diminished the phase transition rapidly completes.
The desired overall effect then is for the curvature coupling to inject a
time-dependence into $\Gamma$, which will act in concert with the time
dependence in $H$ to make $\epsilon$ a more rapidly varying function than
would otherwise be the case and thus suppress the formation of large bubbles.
In the worst case, where $\Gamma$ is approximately constant over the last 60
or so $e$-foldings, we will reproduce the standard extended inflation bubble
spectrum. Any time-dependence in $\Gamma$ helps.

At the point at which the minima become degenerate it can be shown that the
value of the Brans-Dicke field is
\begin{equation}
\Phi_{{\rm deg}}=\frac{24\omega\left(2\omega+1\right)}
	{\left(6\omega+5\right)\left(2\omega+3\right)}
	\frac{\xi \delta M^4}{\gamma^2-4\beta\delta}.
\end{equation}
$\Phi$ enters this calculation through the trapped phase solution for the
scalar curvature.

We restrict the potential in such a way as to ensure that it is always
positive, and denote the zero of the potential by $\chi=\chi_0$. With these
restrictions, we find that for generic parameters the critical value is
similar to the second-order case with $\Phi_{{\rm deg}} \simeq \Phi_c$.
Because we are dealing with more general potentials than in the last section,
the cancellation between the contributions to the present day Planck mass from
the critical $\Phi$ and the vacuum value of $\chi$ is not exact, so a $k$
factor as defined earlier need not be extremely close to unity. In this case
therefore the solar system constraint on $\omega$ for a free $\Phi$ field is
close to the usual one. The question of normalizing the Brans--Dicke field
hinges on three possibilities.
\begin{enumerate}
\item $\Phi_{{\rm deg}}$ is around the value needed to give the correct value
of the Planck mass today. We will assume here that the value of $\xi$ is such
that the subsequent growth in the field $\Phi$ does give us the correct value
of the Planck mass today.
\item A mechanism adjusts $\Phi$ after the phase transition completes.
\item If $\Phi_{{\rm deg}}$ is extremely small one could suppose that bubble
nucleation takes place so slowly that it has time to grow to the value needed
to
produce the correct value of the Planck mass at the point at which the phase
transition completes.
\end{enumerate}

In practice however, the third possibility is of no utility, as by implication
so much inflation would have to occur while $\Phi$ grew that the epoch during
which the curvature term was important would be inflated completely away. It
may also give rise to problems due to the negative $\chi$ contribution to the
Planck mass being greater than the positive $\Phi$ contribution in those
bubbles
that nucleate early. We shall therefore assume that the first of the above
possibilities is realized.

In a typical extended inflation scenario the requirement that the parameter
$\epsilon$ be of order one before there is enough time for `big bubbles' to
grow is found to impose a strong limit of the Brans-Dicke parameter $\omega
\leq 20$ \cite{LW}, which is in disagreement with the limit imposed upon it
by solar system timing experiments. In order to examine the rate of growth of
the bubble nucleation parameter $\epsilon$ in our model, we have used the
tunnelling calculations given by Adams \cite{ADAMS} for a general quartic
potential.

First we can fix parameters using the density perturbation expression
Eq.~(\ref{PERT}), which remains valid, $M$ being as in the second-order case
the false vacuum energy density. As before, large $\omega$ allows us to use
$\Phi_{{\rm deg}}$ in this expression, and assuming field normalization
according to case 1 above we get
\begin{equation}
\label{MEQN}
\frac{M^4}{\xi^2\chi_0^4} = N_1 ,
\end{equation}
where the number $N_1$ is defined to be
\begin{equation}
\label{N1}
N_1 = \frac{300\pi^2\delta_H^2}{\left(\omega+\frac{3}{2}\right)
	\left(\exp\left(\frac{60}{2\omega+1}\right)-1\right)^2} \simeq
	\frac{\pi^2}{3} \, \omega \, \delta_H^2.
\end{equation}
with the last approximate equality holding at large $\omega$.

We make the stringent assumption that we shall only allow 45 $e$-foldings
between the potential being degenerate and $\epsilon$ being of order unity.
The value 45 is chosen because bubbles would have to nucleate before this in
order to be stretched to observationally awkward sizes. If the constraints we
derive seem very strong, it is worth recalling that a much less stringent
condition would also reduce the bubble spectrum on large scales, and further
investigation (hopefully including a discussion of gravitational instantons)
is required before one can be definitive on the merits of our proposal.

We shall be making use here of results from a paper by Adams \cite{ADAMS}. He
provides a numerically derived fitting function for the Euclidean four action,
expressed in terms of the numerical coefficients of the quartic potential,
which is precisely the result we need. Using the usual equation for the decay
probability per unit volume per unit time, which is proportional to the
exponential of the Euclidean four action, and our analytic solution for the
Hubble parameter in the trapped phase, it can be shown that the condition for
the parameter $\epsilon$, as defined above, to be of order one within $45$
$e$-foldings of the minima becoming degenerate is
\begin{equation}
\label{BUBINEQ}
N_2 \left(\frac{\chi_0^4}{M^4} \right) \leq \ln \left(\left( \frac{3}{8\pi}
\right)^2 \frac{m_{Pl}^4}{M^4} \right).
\end{equation}
The numerical factor $N_2$ on the left of this equation depends on the choice
of $\omega$, the number of $e$-foldings between degeneracy of the minima and
$\epsilon=1$, and the geometry of the bare potential. If the geometry of the
bare potential is chosen such that the parameter $\epsilon$ would be of order
one at an early stage during inflation were the curvature coupling not present
(that is, so that the bare $\epsilon$ at the end of inflation would be well in
excess of unity), then the dependance of $N_2$ on the geometry of the bare
potential in the above equation is small. $N_2$ changes significantly when we
change either $\omega$ or the number of allowed $e$-foldings between
degeneracy of the minima and $\epsilon=1$. Assuming $45$ $e$-foldings, then in
the cases $\omega=10000$, $\omega=1000$, $\omega=500$, $\omega=50$ and
$\omega=20$ this numerical factor is found to be $N_2=O(10^5)$, $N_2=400$,
$N_2=80$, $N_2=0.9$ and $N_2=0.07$ respectively.

Assuming the value of the present day Planck mass is obtained as in case 1
above
then it is easily shown that this implies a condition on $\xi$ as follows
\begin{equation}
\label{XIEQ}
\xi=\frac{m_{Pl}^2}{8\pi\chi_0^2\left(\exp\left(\frac{180}{2\omega+1}\right)
	-1\right)} \equiv \frac{N_3 m_{Pl}^2}{\chi_0^2}.
\end{equation}

Using Eqs.~(\ref{MEQN}), (\ref{BUBINEQ}) and (\ref{XIEQ}) the condition for
the phase transition to complete within $45$ $e$-foldings of the tunnelling
rate becoming non-zero, with the correct value of the Planck mass and
reproducing the correct level of density perturbations, is easily shown
to be
\begin{equation}
\label{FININEQ}
\frac{N_2\chi_0^4}{N_1 N_3^2 m_{Pl}^4} \leq \ln \left( \left( \frac{3}{8\pi}
\right)^2 \frac{1}{N_1 N_3^2} \right).
\end{equation}

Given that we have fixed the geometry of the bare potential and have chosen to
allow only $45$ $e$-foldings between degeneracy of the minima and $\epsilon=1$
then the above inequality basically determines the maximum possible value of
$\chi_0$ for any given $\omega$. For each $\omega$ the numbers $N_1$ and $N_3$
are easily evaluated and values of the more complicated number $N_2$ are given
above for some interesting values of $\omega$.

What one finds when this inequality is evaluated for the various values of
$\omega$ given above is that for all $\omega$ in agreement with the
constraint $\omega>500$, all $\chi_0$ values in what we specified to be our
preferred range satisfy the inequality. As $\omega$ is reduced the upper end
of this range is excluded by the inequality.

This appears to be a very interesting result as it implies that the
scenario outlined above allows one to overcome the big bubble problem, even
for large values of $\omega$ which are in agreement with observational
constraints; unfortunately, as in the second-order case the
parameters of the theory are large, taking on their smallest values
at the high end of the preferred $\chi_0$ range. For example, for
$\omega=1000$ and using $\chi_0=10^{-3}$ (which is allowed by the inequality)
then $\xi$ is of order $10^5$ and the quantity $N_1 \xi^2$, which is the
closest equivalent to the coupling $\lambda$ in the second-order case, is also
of order $10^5$. The size of these quantities increases with increasing
$\omega$ (at constant $\chi_0$) and decreases with $\chi_0$ (at constant
$\omega$).

We would like further to point out that in this case introducing a mechanism
to normalize $\Phi$ (and hence the Planck mass) after inflation makes the
constraints harder to satisfy. However, we remind the reader that a detailed
analysis of the bubble constraint would not require such a stringent condition
on the evolution of $\epsilon$, and that any time dependence in $\Gamma$ must
necessarily aid the evasion of the big bubble problem. Only a small
suppression is necessary to enable one to evade the problems associated with
the required flatness of the perturbation spectrum, though should the model
still possess $\omega < 500$ during inflation one would not be able to
dispense with a mechanism such as a mass for $\Phi$ to avoid violation of the
solar system constraints.

In both this and the second-order case, if we demand that the inflationary
perturbations were contributing only part the COBE signal (the remainder
coming say from topological defects) then the couplings could be kept small
for a wider range of $\chi_0$ values.

The results above certainly deserve further detailed examination.

\section{Discussion and Conclusions}

We have investigated models based on extended inflation, with the additional
feature that the inflaton field is coupled to the space-time curvature. This
extension is a particularly natural one, as such a term would be generated by
quantum corrections even if absent in the bare (classical) action.

Several new physical features arise, including
\begin{enumerate}
\item Even with a second-order potential, it becomes possible to have a period
of inflation driven as if by a trapped vacuum phase. Unlike in Einstein
gravity, such a phase ends naturally via an instability induced by the
decreasing spatial curvature.
\item It seems likely that working models exist for $\omega$ chosen large
enough to satisfy the solar system timing constraints with such second-order
models, though a detailed analysis of the rolling phase is essential to
confirm this. Introducing a potential for the Brans--Dicke field introduces
yet more possibilities at the cost of fine-tuning.
\item With an underlying first-order potential, it is possible to construct a
model where inflation ends by bubble nucleation, in a theory with a modified
gravitational sector. This can even be achieved with values of $\omega$
satisfying the solar system bounds without introduction of a potential for the
Brans--Dicke field. Successful completion of the phase transition is brought
about by the inflaton's curvature coupling introducing a time variation into
the nucleation rate $\Gamma$, a mechanism which bears some similarity to
models where modifications of the particle sector alone allow inflation to end
via bubble nucleation \cite{2FIELD}.
\end{enumerate}

We note also that, in common with all extended inflation models, one expects
the production of topological defects at (or close to) the end of inflation
\cite{CKL}. This is because the direction of symmetry breaking in the inflaton
field is only decided at the end of inflation, when the field becomes unstable
to rolling or tunnelling from the origin. (Such an effect is also possible
with two field models of inflation using field couplings and Einstein gravity
\cite{2SCALE}.) The role of topological defects can be very important in
providing extra restrictions on the parameters, and indeed our choice of range
for $\chi_0$ is more or less motivated by this criterion. All types of defects
are problematical if the characteristic scale $\chi_0/m_{Pl}$ is too large.

For the potential as we have been writing it thus far, it would be domain
walls which would form, which would of course be disastrous. All of global
strings, monopoles, textures and nontopological textures exhibit scaling
solutions, and are allowed provided that the energy scale is sufficiently low,
less than around $\chi_0 = 10^{-3} m_{Pl}$ \cite{PST}. In that case all we
have said goes through as before with $\chi$ replaced by $|\chi|$, the angular
degrees playing no role until the instability sets in. One could also imagine
that the inflaton symmetry might be gauged, in which case gauge strings, which
also scale, would not cause any problems if the characteristic scale is below
about the same value \cite{BSB}. Gauge monopoles are rather more interesting,
as they are the only defect whose existence is inevitable if the symmetry is
connected to a Grand Unified Theory. Further, gauge monopoles do not scale,
and the constraints on their formation are much tighter than the other cases,
leading to a much lower limit of $\chi_0 < 10^{-8} m_{Pl}$ or even less
\cite{KT}.

However, even if the values of $\chi_0$ above are violated, it is possible for
consistency with observations to be attained if enough inflation occurs after
the defects form, during the rolling phase. For defects which possess scaling
solutions, one needs to expand the correlation length practically to the
present horizon size, and so one really needs to erase all observational
impact of the trapped phase. For gauge monopoles, however, one can get away
with much less, as one needs only sufficiently dilute their overabundance. The
standard estimates \cite{KT} suggest around $10^{25}$ times too high a number
density at formation, which can thus be remedied by an increase in the
correlation length due to inflation of around $10^8$, corresponding to about
20 $e$-foldings. Models with a free Brans--Dicke field are always very safe,
but those with a potential or based on a first-order potential are capable of
only diluting monopoles to just below observable limits.

In standard inflationary models, the defect constraint is usually weak, as
with standard inflation it is nearly generic for the inflaton scale governed
by large scale structure constraints to be lower than the defect scale
required to have a marked effect on structure formation and microwave
anisotropies \cite{HP}. In this case the restrictions are stronger, because
the Planck mass during the trapped phase is higher than at present, and so the
inflationary scale can be higher.

Of course, according to one's taste, if the symmetries are of a type thought
useful for structure formation (primarily gauge strings or global texture),
then one could try and use them, giving three possibilities
\begin{enumerate}
\item Use the inflaton perturbations for everything, and ensure the defect
scale is too low for any noticeable effects.
\item Use the defects for everything, ensuring that the inflaton perturbations
are too small, to have a noticeable effect.
As stated above, this is rather tricky, but perhaps not
impossible in models with a small $\lambda$ to suppress the vacuum energy with
respect to the expectation value of the true vacuum.
\item Use them both in combination. This is a rather ugly possibility, but
irritatingly it is probably the one best suited to the present data. In
inflation models, the COBE result is generally regarded as being a bit higher
than one would like based on prejudice regarding galaxy formation \cite{LL2}.
Meanwhile, defects are recognized as providing larger microwave anisotropies
for a given size of density perturbation \cite{PST,ALBERT}, but probably to a
greater extent than one would like. In combination, the defects could be used
to soak up some of the excess microwave anisotropies, without having too much
of an effect on the density perturbations.
\end{enumerate}

We have investigated the possibilities introduced by the addition of the
inflaton's curvature coupling. One can use the curvature coupling to
temporarily stabilize the false vacuum even with an underlying second-order
potential. This permits extended inflation to end by classical field evolution
while satisfying the solar-system constraints on $\omega$, though we have not
been able to provide a definitive density perturbation calculation here as the
relevant inflation occurs in the rolling phase rather than the trapped one.
With an underlying first-order potential, it is possible to evade the big
bubble problem which has plagued previous models, and while the overall theory
lacks the simplicity of models such as chaotic inflation, it does have the
advantage of being based on what appears to be fairly plausible particle
physics, and provides a potentially rich phenomenology with topological
defects being formed at the end of inflation. The most likely versions would
however rely on the inflaton field perturbation to provide structure
formation, in which case the generic prediction of a very flat spectrum and
negligible gravitational wave production may be testable in the near future
\cite{GRAV}.

However, there is much scope for further work providing an accurate
delineation of the parameter space of the model, including the development of
a better understanding of density perturbations induced by the rolling phase.
{}From our preliminary numerical studies, the main worry is an unusual one for
inflationary models --- the coupling constants appear to be generically rather
large (in many cases greater than unity) instead of unpopularly small. This
effect has already been noted in the `variable Planck mass' model \cite{SBB},
which has some loose connections with the scenario we have discussed here, and
it would be interesting to discover whether or not this effect is generic.

\section*{Acknowledgements}
The authors are supported by the SERC, and acknowledge the use of the Starlink
computer system at Sussex. We are very grateful to Andrei Linde for sending us
a copy of his preprint \cite{LINP}, and for the many subsequent discussions
which have substantially improved this paper. We also wish to thank David
Wands for discussions concerning conformal transformations, and Jim Lidsey for
his helpful comments.
%%%%%%%%%%%%%%%%%%%%%%%%%%%%%%%%%%%%%%%%%%%%%%%%%%%%%%%%%%%%%%%%%%%%%%

%%%%%%%%%%%%%%%%%%%%%%%%%%%%%%%%%%%%%%%%%%%%%%%%%%%%%%%%%%%%%%%%%%%%%%

\frenchspacing
\begin{thebibliography}{99}
\bibitem{LS} D. La and P. J. Steinhardt, Phys. Rev. Lett. {\bf 62}, 376
	(1989).
\bibitem{K} E. W. Kolb, Physica Scripta {\bf T36}, 199 (1991).
\bibitem{GUTH} A. H. Guth, Phys. Rev. D{\bf 23}, 347 (1981).
\bibitem{KT} E. W. Kolb and M. S. Turner, {\em The Early Universe},
	Addison-Wesley 	(1990).
\bibitem{LINDE} A. D. Linde, Phys. Lett. {\bf 129B}, 177 (1983).\\
	A. D. Linde, {\em Particle Physics and Cosmology}, Harwood Academic
	(Switzerland) (1990).
\bibitem{W} E. Weinberg, Phys. Rev. D{\bf 40}, 3950 (1989).
\bibitem{LSB} D. La, P. J. Steinhardt and E. Bertschinger, Phys. Lett.
	{\bf 231B}, 231 (1989).
\bibitem{RET} R. D. Reasenberg {\it et al}, Astrophys. J. Lett. {\bf 234},
	L219 (1979).
\bibitem{LW} A. R. Liddle and D. Wands, Mon. Not. Roy. astr. Soc. {\bf 253},
	637 (1991).\\ A. R. Liddle and D. Wands, Phys. Lett. {\bf 276B}, 18
	(1992).
\bibitem{AT} F. S. Accetta and J. J. Trester, Phys. Rev. D{\bf 39}, 2854
	(1989).
\bibitem{MOD1} R. Holman, E. W. Kolb and Y. Wang, Phys. Rev. Lett. {\bf 65},
	17 (1990).
\bibitem{MOD2} R. Holman, E. W. Kolb, S. Vadas and Y. Wang, Phys. Rev.
	D{\bf 43}, 3833 (1991).\\ R. Holman, E. W. Kolb, S. Vadas and Y. Wang,
	Phys. Lett. {\bf 269B}, 252 (1991).
\bibitem{LL} A. R. Liddle and D. H. Lyth, Phys. Lett. {\bf 291B}, 391
	(1992).\\ A. R. Liddle and D. H. Lyth, Ann. New York Acad. Sci. {\bf
	688}, 647 (1993).
\bibitem{KST} E. W. Kolb, D. S. Salopek and M. S. Turner, Phys. Rev. D{\bf
	42}, 3925 (1990).\\
	A. H. Guth and B. Jain, Phys. Rev. D{\bf 45}, 426 (1992).\\
	D. H. Lyth and E. D. Stewart, Phys. Lett. {\bf 274B}, 168 (1992).
\bibitem{LL2} A. R. Liddle and D. H. Lyth, Phys. Rep. {\bf 231}, 1 (1993).
\bibitem{COBE} G. F. Smoot {\it et al}, Astrophys. J. Lett. {\bf 396}, L1
	(1992).
\bibitem{SA} P. J. Steinhardt and F. S. Accetta, Phys. Rev. Lett. {\bf 64},
	2740 (1990).
\bibitem{LW2} A. R. Liddle and D. Wands, Phys. Rev. D{\bf 45}, 2665 (1992).
\bibitem{CS} R. Crittenden and P. J. Steinhardt, Phys. Lett. {\bf 293B},
	39 (1992).
\bibitem{LIN82} A. D. Linde, Phys. Lett. {\bf 114B}, 431 (1982).
\bibitem{YOK} J. Yokoyama, Phys. Lett. {\bf 212B}, 273 (1988).\\ J.
	Yokoyama, Phys. Rev. Lett. {\bf 63}, 712 (1989).
\bibitem{CURV} T. Futamase and K. Maeda, Phys. Rev. D{\bf 39}, 399 (1989).\\
        T. Futamase, T. Rothman and R. Matzner, Phys. Rev. D{\bf 39},
        405 (1989).
\bibitem{SBB} D. S. Salopek, J. R. Bond and J. M. Bardeen, Phys. Rev.
	D{\bf 40}, 1743 (1989).
\bibitem{SPOK} B. L. Spokoiny, Phys. Lett. {\bf 147B}, 39 (1984).
\bibitem{SOFT} A. L. Berkin, K. Maeda and J. Yokoyama, Phys. Rev. Lett.
	{\bf 65}, 141 (1990).\\ A. L. Berkin and K. Maeda, Phys. Rev. D{\bf
	44}, 1691 (1991).
\bibitem{BD} C. Brans and R. H. Dicke, Phys. Rev. {\bf 124}, 925 (1961).
\bibitem{LAYCOCK} A. M. Laycock, ``Canonical field redefinition after
        conformal transformations'', in preparation.
\bibitem{LINP} A. D. Linde, ``Hybrid Inflation'', Stanford preprint
	SU-ITP-93-17 (1993).
\bibitem{N} H. Nariai, Prog. Theor. Phys. {\bf 42}, 544 (1969).
\bibitem{DOLGOV} A. D. Dolgov, Phys. Rep. {\bf 222}, 309 (1992).
\bibitem{Lin90} A. D. Linde, Phys. Lett. {\bf 238B}, 160 (1990).
\bibitem{AFG} F. C. Adams, K. Freese and A. H. Guth, Phys. Rev. D{\bf 43},
	965 (1991).
\bibitem{PNP} C. M. Will, {\em Theory and Experiment in Gravitational
	Physics}, Cambridge University Press, Cambridge (1981).
\bibitem{CW} S. Coleman and E. J. Weinberg, Phys. Rev. D{\bf 7}, 1888 (1973).
\bibitem{ADAMS} F. C. Adams, Phys. Rev D{\bf 48}, 2800 (1993).
\bibitem{2FIELD} A. D. Linde, Phys. Lett. {\bf 249B}, 18 (1990).\\
	F. C. Adams and K. Freese, Phys. Rev. D{\bf 43}, 353 (1991).
\bibitem{CKL} E. J. Copeland, E. W. Kolb and A. R. Liddle, Phys. Rev. D{\bf
	42}, 2911 (1990).
\bibitem{2SCALE} A. D. Linde, Phys. Lett. {\bf 259B}, 38 (1991).
\bibitem{PST} U.-L. Pen, D. N. Spergel and N. Turok, ``Cosmic Structure
	Formation and Microwave Anisotropies from Global Field Ordering'',
	Imperial College preprint Imperial/TP/92-93/04 (1992).
\bibitem{BSB} D. P. Bennett, A. Stebbins and F. R. Bouchet, Astrophys. J.
	Lett. {\bf 399}, L5 (1992).
\bibitem{HP} H. M. Hodges and J. Primack, Phys. Rev. D{\bf 43}, 3155 (1990).
\bibitem{ALBERT} A. Stebbins, Ann. New York Acad. Sci. {\bf 688}, 824 (1993).
\bibitem{GRAV}	E. J. Copeland, E. W. Kolb, A. R. Liddle and J. E. Lidsey,
	Phys. Rev. Lett. {\bf 71}, 219 (1993).\\ R. Crittenden, J. R. Bond, R.
	L. Davis, G. Efstathiou, and P. J. Steinhardt, Phys. Rev. Lett.
	{\bf 71}, 324 (1993).\\ E. J. Copeland, E. W. Kolb, A. R. Liddle and
	J. E. Lidsey, Phys. Rev. D{\bf 48}, 2529 (1993).
\end{thebibliography}
\end{document}